\documentclass{aa501}
\usepackage[]{natbib}
\usepackage[]{graphics}
\bibpunct{(}{)}{;}{a}{}{,}
\newcommand{\etal}{{\it et al.}\ }
\newcommand{\Msun}{\hbox{M$_\odot$}}
\newcommand{\sbu}{\mbox{mag arcsec$^{-2}$}\ }
\newcommand{\kms}{\mbox{km~s$^{-1}$}}
\begin{document}

\title{The gas content of peculiar galaxies: counterrotators and polar 
rings\thanks{Based on 
observations collected at SEST telescope, European Southern Observatory, La 
Silla, Chile.} \fnmsep
\thanks{Table 1 is only available in electronic form.} }

\author{D. Bettoni\inst{1}
\and
G. Galletta\inst{2}  
\and
S. Garc\'{\i}a-Burillo\inst{3} 
\and
A. Rodr\'{\i}guez-Franco\inst{4,5}}
   \offprints{G. Galletta \\ \email{galletta@pd.astro.it}}

\institute{Osservatorio Astronomico di Padova, Vicolo Osservatorio 5, 35122 
Padova, Italy
\and
Dipartimento di Astronomia, Universit\`a di Padova, Vicolo Osservatorio 2, 
35122 Padova, Italy
\and
Observatorio Astron\'omico Nacional-OAN, Apartado 1143, 28800 Alcal\'a de 
Henares- Madrid, Spain
\and
Departamento de Matem\'atica Aplicada (Biomatem\'atica) Secci\'on 
Departamental de Optica. Universidad Complutense de Madrid, Av. Arcos de 
Jal\'on s/n, 28037 Madrid, Spain.
\and
Nobeyama Radio Observatory, Nobeyama, Minamimaki, Minamisaku, Nagano, 
384-1305, Japan.}

\date{Received 23 February 2201/Accepted 7 May 2001}

\authorrunning{Bettoni et al.}
\titlerunning{Gas content of peculiar galaxies}

\abstract{
This paper studies the global ISM content in a sample of 104 accreting
galaxies, including counterrotators and polar rings, which spans the
entire Hubble sequence. The molecular, atomic and hot gas content of
accretors is compared to a newly compiled sample of normal
galaxies. \\ 
We present results of a small survey of the J=1--0
line of $^{12}$CO with the 15m SEST telescope on a sample of 11
accretors (10 counterrotators and 1 polar ring). The SEST sample is
enlarged with published data from 48 galaxies, for which observational
evidence of counterrotation in the gas and/or the stars has been
found. Furthermore, the available data on a sample of 46 polar ring
galaxies has been compiled. In order to explore the existence of an
evolutionary path linking the two families of accretors, the gas
content of counterrotators and polar rings is compared. \\ 
It was found that the normalized content of cold gas (M$_{gas}$/L$_B$) in
polar rings is $\sim$1 order of magnitude higher than the reference
value derived for normal galaxies. The inferred gas masses are
sufficient to stabilize polar rings through self-gravity. In contrast,
it was found that the cold gas content of counterrotators is close to
normal for all galaxy types. Although counterrotators and polar rings
probably share a common origin, the gas masses estimated here confirm
that {\it light} gas rings accreted by future counterrotators may have
evolved faster than the self-gravitating structures of polar rings. In
this scenario, the transformation of atomic into molecular gas could
be enhanced near the transition region between the prograde and the
retrograde disks, especially in late-type accretors characterized by a
high content of primordial gas. This is tentatively confirmed in this
work: the measured H$_2$/HI ratio seems larger in counterrotators than
in normal or polar ring galaxies for types later than S0s.
 }

\maketitle

\section{Introduction}\label{intro}
   
The existence of kinematically decoupled disks of gas and/or stars
with anti-parallel spins has been reported for a significant number of
galaxies \citep[see ][ for a review]{gall96}. The phenomenon of
counterrotation may be seen in the ionized gas \citep{bettoni84} but
in almost half of the reported cases it is found in pure {\it stellar}
disks \citep[see ][ for a review]{rubin}, being in some cases
accompanied by gas counterrotation. Evidence of kinematical decoupling
for the cold gas, either atomic or molecular, is also present in a
high percentage of counterrotating galaxies \citep{oosterloo, braun,
casoli, vandriel, sage2, santi, santi2}.

Different scenarios have been proposed to explain the counterrotation
present in elliptical and disk galaxies. Most of them invoke the
capture of matter which comes from outside the acceptor galaxies. The
various models examine different masses and time-scales involved
in the accretion process. An external origin is also invoked to
explain the existence of polar ring galaxies, where gaseous disks or
rings are seen to rotate almost perpendicularly with respect to the
main stellar body of the system \citep{whitmore}. However, the link
between polar rings and counterrotators remains unclear. Alternatively, 
it has been suggested that a primordial mechanism, invoking a dissipationless 
cosmological collapse perturbed by tidal fields, could explain the formation 
of counterrotating galaxies \citep{harsoula}.

Within the accretion scenario, the morphology of the acceptor system
and its dynamic evolution would depend on several factors: the nature
of the accreted matter (gas and/or stars), the ratio between the
accreted mass and that of the acceptor galaxy and, finally, the
accretion speed. Whereas the collision between equally massive
galaxies may lead to a merging like the `Antennae', ending up as a
giant elliptical galaxy with counterrotation
\citep{barnes}, the disruption of the acceptor´s disk could be avoided by 
progressive infall of gas whose spin is anticorrelated with the main
stellar body \citep{quinn,thakar,voglis}. However, the accretion of a
gas-rich satellite may heat the stellar disk
\citep{thakar}. Observations show that lenticulars and spiral galaxies
hosting counterrotation do not necessarily present disrupted stellar
disks. Their stellar kinematics appear globally regular
\citep{oosterloo} and the stellar disks show hardly any sign of
thickening \citep{n4550}. The end product of the accretion process at
the present epoch seems to have reached, in most of the known studied
cases, an equilibrium configuration for the stellar component. Either
the time-scales to reach equilibrium are short enough or,
alternatively, accretion caused no traumatic changes in the kinematics
of the stars.

The gas, however, is expected to reflect the consequences of the
accretion process more violently than the stars. If gas was accreted
by a disk galaxy with a non negligible amount of interstellar gas, a
strong interaction between the accreted and the primary gas is
likely. The existence of violent cloud-cloud collisions (the relative
velocity between the interacting clouds would be: v$\sim$2\,v$_{rot}$)
and the highly dissipative nature of the encounters might lead to the
onset of large-scale shocks. These might convert the atomic gas into
molecular gas \citep{braine,sage1,young} and eventually induce
starbursts \citep{wang, read, santi2}. If the described scenario
holds, one would expect that the content of molecular gas would be
higher in counterrotating galaxies than in a comparison sample of
non-interacting galaxies of the same Hubble type. On the other hand,
if the origin of counterrotation is primordial, or alternatively, if
large-scale shocks are not efficient or short-lived, the H$_2$ content
should be similar for counterrotating and normal galaxies.

It is also unclear whether polar rings and counterrotators represent
different steps in the process of mass accretion. A comparison of
their H$_2$ content could reveal if there is an evolutionary link
between the two families of accretors.

This paper represents a first step in answering some of the
above-mentioned questions by studying the global gas content in a
sample of counterrotators.  In this work we estimates the content of
molecular, atomic and hot gas for a sample of 58 galaxies of different
morphological Hubble types and of different types of counterrotation. 
Molecular gas masses are derived from $^{12}$CO(J=1--0) 
observations made with the 15m SEST radiotelescope on 10 galaxies with
counterrotation and 1 polar ring (section 2).  Results from these new
observations are described in Section 4. The published data for 48
objects where there are indications of counterrotation in the gas
and/or stars have been compiled and added to this sample(see Section
5).  The H$_2$ content of counterrotating galaxies will be studied
relative to a comparison sample of normal galaxies that was built
up from the literature, as described in section 6. The gas content of
counterrotators has also been compared with those of polar ring
galaxies, using available data from the literature (see below for
detailed references). A similar comparative study has been done with
the HI content (from various sources), with the warm dust mass
(derived from IRAS data) and with the amount of hot gas, the dominant
gas component in early type objects (values derived from X-ray data
taken by ROSAT and by EINSTEIN).

\section{The SEST sample}

The 11 galaxies in our sample observed using the SEST telescope were
selected from the list of objects published by \citet{gall96}.  The
galaxies are described individually in section \ref{results}, together
with the main results inferred from this CO study.  The relevant
parameters of the systems, such as the diameters (D$_{25}$: the
de-projected linear diameter corresponding to the blue isophote at 25
\sbu), absolute B magnitudes (M$_B$), distances ($d$) and
morphological types are shown in columns 3--6 of Table 1. 
These data have been extracted from the recently
up-dated Lyon-Meudon database LEDA \citep{leda}.

The conversion factor between the integrated intensities of
$^{12}$CO(J=1--0) and the H$_2$ column densities was taken from
\citet{strong}, i.e.:

$$\chi=N(H_2)/I_{10}=2.3\times10^{20} mol/Kkms^{-1}. $$

The total mass of molecular hydrogen (in \Msun) under the 45\arcsec\
beam of SEST for the observed galaxies was obtained from N(H$_2$)
using: $$ M(H_2)=7.8\, 10^{-16}\times N(H_2)\times d^{2} (\Msun)$$
where $d$ is the galaxy distance in Mpc and N(H$_2$) the measured
column density in mol\,cm$^{-2}$. The total molecular gas content
under the beam (M$_{mol}$) is derived by multiplying M(H$_2$) by 1.36
in order to include the Helium mass fraction. Although a variation of
the $\chi$ conversion factor cannot be excluded among the observed
galaxies, we will take the above derived values as a good estimate of
the molecular gas masses. 

HI masses (M$_{HI}$) have been taken from various sources or
calculated from the m$_{21}$ parameter of LEDA \citep{leda}, assuming:
$$M_{HI} = 2.35\ 10^5\ 10^{-0.4(m_{21c}-17.4)}\ d^2 $$ with $d$ being
the distance in Mpc. Hereafter, M$_{gas}$ is used for the total mass
of cold gas, i.e., M$_{gas}$=M$_{mol}$+M$_{HI}$.

The mass values of X-ray emitting gas (M$_X$) have been derived from
ROSAT data \citep{beuing} and from EINSTEIN data \citep{fabbiano,
burstein}, assuming: $$M_{X} = 10^{-24}\ L_x^{0.5}\ L_B^{1.2} $$ Note
that this formula \citep{roberts} is generally valid for early-type
galaxies.

Finally, warm dust masses (M$_{dust}$) have been calculated from IR
fluxes (S$_{60}$ and S$_{100}$) published by \citet{knapp} and from
LEDA raw data, kindly furnished by G. Paturel, assuming: $$M_d = 4.78\
10^{-3}\ S_{100}\ d^2\ (exp(144.06/T_d)-1) $$ where dust temperature
$T_d=49*(S_{60}/S_{100})^{0.4}$ and d is the distance in Mpc. A more
accurate calculation giving the total dust mass, including the coldest
component not detected by IRAS, is beyond the scope of this work.
 
All distance-dependent values available from the literature have been
re-scaled to the adopted distances from LEDA, explicitly listed in
Table 1.

\section{Observations}\label{obs}

Emission in the $J$=1--0 and $J$=2--1 transitions of $^{12}$CO among
the sample of 11 galaxies was searched for using the Swedish-ESO
Submillimeter Telescope (SEST) at La Silla, equipped with the dual
channel IRAM 115/230 GHz receivers which allow for simultaneous
observations. There were three different observation runs: November
6th-11th 1998, May 28th--31st 1999 and May7th-11th 2000. Beam sizes
were 45\arcsec\ and 23\arcsec\ at 115 GHz and 230 GHz,
respectively. Unless explicitly stated, the temperature scale used
throughout the paper is antenna temperature, corrected from
atmospheric losses and rear spillover (T$_a^*$). When deriving line
ratios, it is assumed that main beam efficiencies $\eta_{beam}$(115
GHz)=0.70 and $\eta_{beam}$(230 GHz)=0.50, in order to refer
temperatures to the main beam brightness scale.  Spectrometers cover
bandwidths of 995 MHz (1290 \kms) and 543 MHz (1410\kms) for the J=2--1
and J=1--0 lines respectively. Dual-beam switching was used, with a beam
throw of 12\arcmin\ to produce a flat baseline. Typical system
temperatures ranged from 190-430 K. Pointing and focus were checked
every 2-3 hours, using several SiO maser sources located near the
target galaxy. The RMS accuracy of the pointing model was typically
$\le$2\arcsec, assuring an absolute positional accuracy better than
5\arcsec.

Except for one case (NGC\,3497), we made only single point maps
centered on the nuclei of the galaxies. Individual scans at each
position were coadded to get total integration times ranging from 1h
to 7h.  Spectra were Hanning-smoothed to a velocity resolution of
30 \kms\ (for both lines) with the exception of narrow lines for which
a higher resolution was kept (see below). Linear baselines were fitted
and subtracted from the smoothed spectra using the GILDAS software
package.

\begin{figure*}
\vskip -2cm
\resizebox{\hsize}{!}{\includegraphics{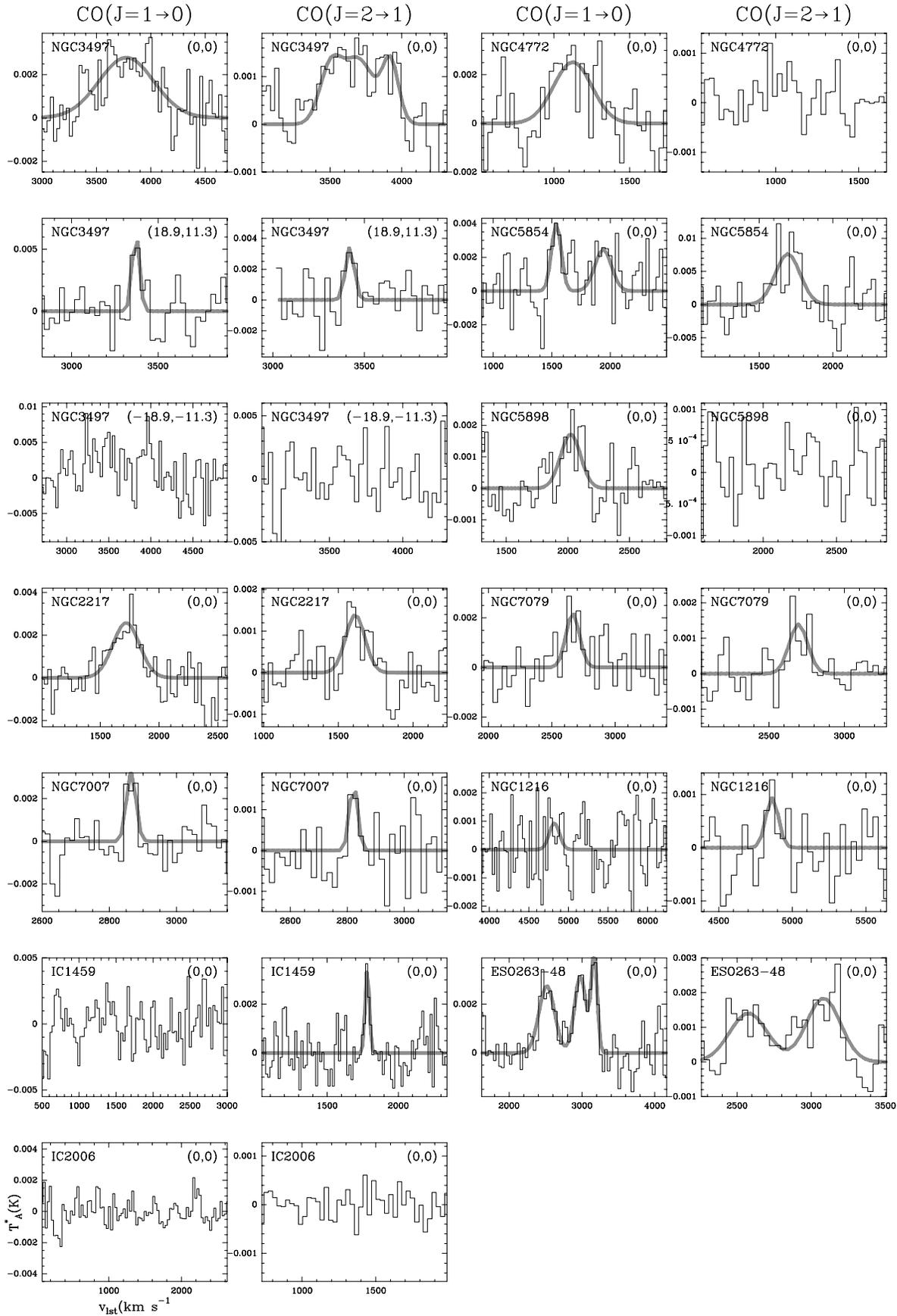}}
\caption{The J=1--0 and J=2--1 CO spectra for the galaxies observed by the SEST 
telescope. Gaussian fits on the lines are shown when available.}
\label{CO}
\end{figure*}

\section{Results for the SEST sample}\label{results}

This section presents the main results of the CO observations for the 11 
galaxies of the SEST sample, preceded by a short description of the systems.

\subsection{ESO263-48}

This galaxy, also denoted Anon 1029-45, has been included in a list
of dust-lane ellipticals by \citet{hawarden}. It has a prominent and
strongly warped dust lane going across the major axis up to
r=30\arcsec\ (5.2 kpc). Due to its prominent dust lane which gives the
system the appearance of an edge-on disk, it has often been classified
as S0, although it has all the properties (photometric profile, luminosity
and size) of giant ellipticals. Furthermore, long exposure plates of
the galaxy show no signature of a stellar disk.
 
The stellar kinematics have been studied by \citet{bertola1}, who
derived a maximum rotational velocity of 210 \kms\ reached at
r=20\arcsec\ (3.4 kpc) and a velocity dispersion of 260 \kms. Ionized
gas in counterrotation, with spin velocities of $\sim$ 250 \kms\ at
r=7\arcsec\ (1.2 kpc), has been detected along the major axis
\citep{bertola2}. There are no HI observations available for
ESO263-48; the galaxy also remains undetected in the ROSAT
survey. However, this elliptical is particularly rich in dust; from
the IRAS fluxes we estimate M$_{dust} \sim$3.8\,10$^{5}$\Msun. A
continuum radio emission has been detected at 1.4 and 4.9 GHz, which
extends perpendicularly to the dust lane \citep{bertola1}.
 
The J=2--1 and J=1--0 CO spectra of Fig.1 show $\sim$800 \kms--wide
emission profiles centered at v=2810 \kms\ (hereafter taken as the
CO-based systemic velocity; v$_{sys}$). The J=1--0 spectrum is
asymmetrical with respect to v$_{sys}$ as it can be fitted by three
gaussian-like components. Emission of the two extreme velocity
components at v=2517 \kms\ and v=3170 \kms\ can be explained by the
presence of an unresolved H$_2$ disk with a rotation speed of
v$_{rot}$=325 \kms, reached within r$=$2 kpc (the upper limit on r is
set by the J=2--1 beam). The value of v$_{rot}$ derived from CO is
significantly larger than that which is inferred for the ionized gas
at the same radius ($\sim$250 \kms). This discrepancy may indicate
that the H$_2$ disk seen in projection extends farther out. The
asymmetry in the $^{12}$CO(J=1--0) spectrum is caused by the existence
of a strong component at V=2980 \kms, redshifted by 170 \kms\
with respect to v$_{sys}$, and having no blue-shifted counterpart. The
latter can come from an asymmetrical distribution in the H$_2$ disk
or, alternatively, be the signature of a warp in the molecular gas
disk (as suggested by the distorted dust lane).

The molecular gas content under the J=1--0 beam can be derived
within a radius r=22\arcsec\ (4 kpc) (close to the maximum extent of
the dust lane feature). It was calculated to be M$_{mol}$=8\,10$^{8}$
\Msun, a high value for an elliptical galaxy.

\subsection{IC\,1459}

The absolute magnitude and the intrinsic diameter of this galaxy (see Table 
1) are both characteristic of a giant elliptical. IC\,1459 has a 
massive counterrotating stellar core (M$\sim$10$^{10}$ \Msun, according to 
\citet{franx}) and its outer stellar isophotes are twisted, an indication that 
the stellar body is triaxial. This galaxy shows dust absorption in the central 
10\arcsec\ \citep{sparks} and faint pseudo-arms in the outer part 
(r$\sim$3.5\arcmin\, \citet {malin}). The counterrotating core, which hosts a 
compact radio source \citep{franx}, has a radius of $\sim$2\arcsec\  (200 pc) 
and a projected spin velocity of 170$\pm$20 \kms. On dynamical grounds, the 
core can be described as a disk, rather than as an ellipsoid. In contrast, the 
outer directly-rotating stellar body has a slower rotation figure; it reaches 
$\sim$45$\pm$8 \kms\ at r=40\arcsec\ (4 kpc). The galaxy is crossed by a disk 
of ionized gas, whose emission is evenly detected up to r=35\arcsec\ (3.5 
kpc); the ionized gas rotates in the same direction than the outer stellar 
body, but at a higher speed (350 \kms\ ) \citep{franx}. Therefore 
counterrotation in this galaxy seems confined to the inner core and affects 
only the stars. X-ray emission peaks in the galaxy nucleus \citep{roberts}, 
an indication of central activity.

\citet{walsh} report a negative detection of HI emission,  
the upper limit on M$_{HI}$ being $\sim$10$^7$ \Msun. Our $^{12}$CO(J=1--0)
spectrum also shows a negative detection for molecular gas emission.  In
contrast, the integrated $^{12}$CO(J=2--1) spectrum shows clearly a narrow
emission line of FWHM$\sim$35 \kms, centered at 1782 \kms\  (see Fig.1). 
The velocity centroid of CO is redshifted by $\sim$100 \kms\  with respect 
to the galaxy systemic velocity (v$_{sys}$=1691 \kms).  This discrepancy 
of velocity centroids  and the narrowness of the $^{12}$CO(J=2--1) line 
both indicate that emission cannot come from a rotating
molecular gas disk in equilibrium that could be associated with the ionized
disk or, alternatively, with the counterrotating core. In the discarded
scenario of a molecular gas disk, the CO line should be 3--5 times wider than
is actually observed, considering the size of our beam. Instead, the CO
profile may come from a Giant Molecular  Association (GMA).  Furthermore, the
derived upper limit on the mass of molecular gas, M$_{mol}<$2 10$^7$
\Msun, is noticeably low. Additional support for the interpretation outlined
above comes from the high resolution deconvolved V-band images of the dust
distribution, obtained with the HST. The morphology of the dust lane source
in the inner  4\arcsec\  of the galaxy is very irregular and indicates
non equilibrium motions \citep{forbes}. Finally, the estimated 3$\sigma$ upper
limit on the (J=2--1)/(J=1--0) ratio ($>$1.3)  suggests that the CO emission might be
partly optically thin. 

Similar molecular gas components have been found in other
early-type galaxies, such as NGC\,404 \citep{wiklind}, classified as a gas
accreting elliptical with a minor axis dust lane \citep{bg}. This GMA
may be a residual of one of the galaxies  involved in the passed merger
that is supposed to be at the origin of IC\,1459 \citep{franx}.

\subsection{IC\,2006}

The size and the luminosity of IC 2006 put this galaxy among the dwarf
ellipticals. Counterrotation is present in a outer ring of atomic gas,
which is aligned with the apparent major axis of the galaxy. At the
radius of the HI ring, the galaxy luminosity decreases to B$\sim$27
\sbu \citep{schweizer}.  Schweizer and collaborators describe the HI
distribution as a 2\, kpc wide circular ring of radius $\sim$\,11 kpc,
inclined at 37\degr. With the adopted parameters, the rotation speed
of the HI gas would be 200 \kms\ at this distance. HI gas remains
undetected inside the ring. In contrast, faint emission from ionized
gas is detected within 2.5\, kpc of the nucleus, characterized by a
velocity gradient which is smaller and also inverted with respect to
that of the stars (from -70 \kms\ (NE) to 50 \kms\ (SW), relative to
v$_{sys}$=1385 \kms). The counterrotating ionized gas disk is highly
turbulent, with a measured velocity dispersion of 190 \kms\
\citep{schweizer}.

The upper limit set by ROSAT observations \citep{beuing} indicates
that, contrary to IC\,1459, IC\,2006 has no relevant quantity of hot
gas (see Table 1). Based on the optical photometry and the
kinematics of the outer HI ring, \citet{schweizer} derive the presence
of a dark halo which contains about twice the mass of the luminous
stellar body. Our single-point $^{12}$CO(J=1--0) map sampled the galaxy
nucleus up to r=3.5 kpc. This region includes the ionized gas disk but
excludes the outer HI ring. CO emission was not detected; the latter
implies an upper limit for the central (r$<$3.5 kpc) molecular gas
content of M$_{mol}<$1.4 10$^7$ \Msun.

\subsection{NGC\,1216}

This galaxy belongs to the Hickson compact group N\,23. It is an
almost edge-on disk galaxy classified as S0-a in the LEDA database.
Counterrotation in the ionized gas is detected from optical emission
lines \citep{rubin2}. The gaseous disk extends up to a radius of 2
kpc, with an observed maximum rotational speed of 75 \kms; this
velocity is significantly lower than that of the stars at the same
radius ($\sim$175 \kms).  The latter may be an indication that the
gaseous disk is either warped or tilted with respect to the galaxy
plane.  The optical images of this galaxy show no signature of dust
absorption.  \citet{williams} report a negative detection for this
galaxy in the 21cm line, which gives an upper limit of
M$_{HI}$$<$5\,10$^{7}$\Msun. IRAS, ROSAT and, finally, our CO
observations report negative detections for NGC\,1216. The latter give
an upper limit of M$_{mol}<$3\,10$^8$ \Msun for the molecular gas
content, within a region of radius r=7 kpc.

 \subsection{NGC\,2217}

NGC\,2217 is a barred spiral (SBa) seen almost face-on. It has an
outer stellar+gaseous ring of radius $\sim$8\, kpc and an inner oval
ring of radius $\sim$4\, kpc, which encircles the bar.

\citet{bettoni90} have studied the distribution and kinematics of the ionized
gas, confined to the innermost 20\arcsec\ ($\sim$2 kpc) of the
galaxy. The gas distribution suggests the presence of a two-arm spiral
structure, whereas the kinematics are characterised by counterrotation
with respect to the stars, inside r$\sim$10\arcsec\ ($\sim$1
kpc). However, a detailed analysis of the data by the authors shows
that the gas counterrotation is not real, and it may be better
accounted for if one assumes the gas to be in a warped disk seen in
projection. The ionized gas inside r=1 kpc would lie in a series of
polar rings almost at 90\degr with respect to both the bar and the
stellar disk. In the outer region (r=1-2 kpc) the plane of the gas
rings would have settled towards the disk of the stars, and changed
its inclination by nearly 90 degrees. The latter explains why the gas
and the stars rotate in the same direction for r$>$1 kpc. We have
classified the system as a polar ring galaxy.

The apparent counterrotation of ionized gas inside r=1 kpc produces the 
largest extention of radial velocities along the bar minor axis: from v=1450 
\kms\ to 1800 \kms. The v$_{sys}$ derived from the gas and the star 
kinematics agree within the margin of errors, being close to 1640 \kms. 
\citet{bettoni90} fit a rotation curve to their data, inferring de-projected 
rotational velocities of 125 \kms\ and 150 \kms\ for the stars and the gas, 
respectively, at a radius r=1 kpc. 

HI observations reported by \citet{h82} show a double-horned emission
profile centered at v$_{sys}$=1615 \kms, close to the value found by
\citet{bettoni90} and with a total width at zero power of $\sim$300
\kms. The re-scaled HI mass is M$_{HI}$=2.7\,10$^{9}$\Msun. As there
is no high-resolution HI map of NGC\, 2217, the location of the HI gas
is uncertain.

The spectra in Fig.1 show the detection of molecular gas emission for
the two lines of $^{12}$CO. The two profiles differ significantly,
however.  Whereas the J=1--0 line is centered at v=1720 \kms\ (i.e.,
redshifted 100 \kms with respect to v$_{sys}$), the J=2--1 line peaks at
v$\sim$v$_{sys}$ (as defined above). Although the low spatial
resolution of these observations tells us little on the precise
location of molecular clouds, the reported asymmetry of the J=1--0
profile (which samples the disk up to r=2 kpc) suggests that the
distribution of H$_2$ gas in NGC\,2217 is highly asymmetrical and/or
that the kinematics of molecular clouds might depart from circular
motions. Most noticeably, the integrated HI profile also shows a
pronounced asymmetry. The FWHM of the two CO lines are close to the
values found in HI and in optical emission lines ($\sim$250-300 \kms).

We derived a molecular gas content of M$_{mol}\sim$9 \,10$^{7}$\Msun
up to r=2 kpc.

\subsection{NGC\,3497}

NGC\,3497 is a major-axis dust lane elliptical known by different
names (NGC\,3525=\-NGC\,3528=\-IC\,2624). The ringed dusty disk shown
in the B-R color maps of \citet{eb85} seems to have the same extent
as the stellar disk (diameter$\sim$70\arcsec).  As with all major-axis
dust-lane ellipticals, e.g. ESO 263-48 in this paper, the dust
signature is interpreted to be the result of an accretion episode.
The galaxy has a fainter galaxy at 2\arcmin\ (named NPM1G -19.0362)
and a companion with similar redshift at 5\arcmin\ (NGC 3529=IC2625).

Stellar and gas rotation curves have been derived by \citet{bertola1},
who measured a radial velocity difference of $\sim$240$\pm$30 \kms\
between the western and eastern sides of the major axis (on the NE
side stars are receding). The measured systemic velocity is
v$_{sys}$=3672$\pm$25 \kms.  No X-ray, IR or HI data are available in
the literature.

The emission of both lines of $^{12}$CO were observed in three
positions located along the major axis of the disk: the (0,0) offset
centered on the galaxy nucleus and two off-centered positions at
r=$\pm$22 \arcsec. The J=2--1 and J=1--0 spectra shown in Fig.1 reveal the
presence of molecular gas in the central region (up to r=5 kpc) and
also the detection of the J=1--0 line of $^{12}$CO in the NE offset. In
the SW position, however, CO emission was not detected. The
$^{12}$CO(J=1--0) central spectrum is fitted well by a single gaussian 
profile, centered at 3774 \kms\ and with FWHM=600 \kms\ ($\sim$1000
\kms\ at zero power); therefore, it is 100\kms redshifted with respect to 
v$_{sys}$. In contrast, the J=2--1 profile shows three velocity components at 
v=3497 \kms, v=3693 \kms\ (close to the optically determined v$_{sys}$) and 
v=3497 \kms. The velocity asymmetry in the central J=1--0 spectrum may indicate 
that H$_2$ distribution is slightly asymmetrical in the disk within r=5 kpc. 

A comparison between the radial velocity measured on the CO spectrum
($\sim$3570 \kms) and the stellar velocities observed in the NE side
of the major axis (redshifted with respect to v$_{sys}$), indicates
that molecular gas is counterrotating with respect to the stars.

The molecular gas mass within the central r=5 kpc, derived from the
$^{12}$CO(J=1--0) integrated intensity, would be 
M$_{mol}$=1.4\,10$^{9}$\Msun.  The amount of molecular gas detected in
the NE offset is M$_{mol}$=2.7\,10$^{8}$\Msun.

\subsection{NGC\,4772}

\citet{haynes} considers NGC\,4772 as a case of apparent counterrotation of 
the ionized gas versus the stars. The kinematical decoupling of the
nuclear ionized gas (r$<$5\arcsec=0.3 kpc), observed along both the
minor and the major axes, has been interpreted as the signature of a
misaligned embedded gas bar, rather than as evidence of
counterrotation.  However, this Sa galaxy shares many features with
other prototypical counterrotators. Mimicking NGC\,3626, the HI
content of NGC\,4772 is distributed in two separate rings, probably
non coplanar. As is the case for NGC\, 3626, the central region of
NGC\,4772 is HI-poor. Furthermore, the deep optical photometry
of the galaxy reveals the presence of a round, low surface brightness
disk in the outer part, reminiscent of a similar feature reported by
\citet{buta} in the Sab counterrotator NGC\, 7217.

The J=1--0 spectrum of $^{12}$CO (Fig.1) reveals the presence of
molecular gas (inside r=1.5 kpc). The line profile, centered at
$\sim$1120 \kms\ and with FWHM$\sim$300 \kms, is slightly redshifted
with respect to the optically determined v$_{sys}$=1040 \kms (the same
as derived from HI).

Emission in the J=2--1 line is undetected, however.  The molecular gas
mass within the central r=1.5 kpc, derived from the $^{12}$CO(J=1--0)
integrated intensity, would be M$_{mol}$=5.4\,10$^{7}$M$_{\sun}$.

\subsection{NGC\,5854}

NGC\,5854 is an early spiral (Sa) characterized by a low gas content
and the absence of current star formation. \citet{haynes} have studied
the stellar kinematics, using H$\alpha$ and MgIb optical absorption
lines, and the kinematics of ionized gas, using the N[II] and O[III]
optical emission lines. These data reveal the existence of a
counterrotating gas disk extending up to r$\sim$7\arcsec\ (0.8 kpc),
with a total velocity range of 120 \kms. The stellar velocities
measured at r$\sim$40\arcsec\ (4.5 kpc) reach $\pm$160 \kms. Although
HI content is low, \citet{magri} detected a signal in the nucleus. The
HI profile is centered at 1663 \kms, close to the optically determined
value for v$_{sys}$=1669$\pm$30 \kms\ \citep{fouque}. The narrowness
of the HI spectrum (FWHM$\sim$100 \kms, namely, less than the measured
stellar velocity spread) suggests a close association of the HI
component with the counterrotating ionized gas (see discussion in
\citet{haynes}).

Although weak, the $^{12}$CO(J=1--0) spectrum shows the existence of
H$_2$ gas within r=2.3 kpc. There is a hint of a double-horned
profile, with two velocity components at v=1539 \kms\ (with
FWHM$\sim$100 \kms) and v=1930 \kms\ (with FWHM$\sim$170 \kms),
equidistant from v=1735 \kms. The CO spectrum in the J=2--1 line
confirms the detection of molecular gas. Not surprisingly, the J=2--1
and J=1--0 profiles differ. The J=2--1 line shows hints of two velocity
components, although with a smaller velocity spread (v=1630 \kms\ and
1740 \kms) and is centered at v$\sim$1690 \kms, in reasonable agreement
with HI. However the larger velocity spread of the J=1--0 spectrum would
suggest that, compared to the counterrotating ionized gas core, the
H$_2$ disk may extend farther out.  The molecular gas mass within the
central r=2.3 kpc was derived from the $^{12}$CO(J=1--0) integrated
intensity giving; M$_{mol}$=1.6\,10$^{8}$M$_{\sun}$.

\subsection{NGC\,5898}

NGC\,5898 was studied by \citet{bettoni84} and \citet{bertola88}, who
discovered the first case of ionized gas counterrotation in a dust
lane elliptical in this galaxy. Their data, which extended out to
$\sim$10\arcsec\ (1.4 kpc), have recently been completed by
\citet{caon} who analysed the stellar and the gas kinematics farther
out (up to r$\sim$35\arcsec\ (4.8 kpc)). The new data along the major
axis show the existence of a stellar core of radius r$\sim$5\arcsec\
(0.7 kpc) which counterrotates with respect to the outer stellar
body. The ionized gas counterrotates with respect to the inner stellar
core, and it therefore corotates with the outer stars. In contrast,
gas is seen to counterrotate at all radii along the minor axis. This
might indicate that angular momentum vectors of the ionized gas and
the stars are certainly misaligned, but not antiparallel.  The
observations of this galaxy at X-ray and IR wavelengths show an upper
limit for hot gas of $\sim$3\,10$^8$ \Msun and a moderate quantity of
dust, of a few 10$^4$ \Msun.

The J=1--0 spectrum of $^{12}$CO (Fig.1) shows a tentative detection of
molecular gas inside r=3.1 kpc. The line profile, centered at
$\sim$2020 \kms and with FWHM$\sim$190 \kms, is slightly blueshifted
with respect to the optically determined value of v$_{sys} (\sim$2100
 \kms, derived from
\citet{bertola88}). The observed asymmetry might arise if the H$_2$ gas was 
associated with the ionized disk, which shows a marked extension towards the 
SW (where ionized gas radial velocities are blueshifted). In this scenario 
molecular gas would also be counterrotating. 

The $^{12}$CO emission is undetected in the J=2--1 line, however. We have
calculated the molecular gas mass within the central r=3.1 kpc, using the
$^{12}$CO(J=1--0) integrated intensity, giving M$_{mol}$=10$^{8}$M$_{\sun}$.

\subsection{NGC\,7007}

The optical images of NGC 7007 show an elliptical-like body encircled
by an off-centered bow-shaped dust lane on the eastern
side. \citet{dettmar} discovered, in this galaxy, the signature of a
counterrotating ionized gas disk by comparing the spectrograms of gas
emission (NII $\lambda$6853) and stellar absorption lines
(H$_{\alpha}$). Spectra taken later \citep{bettoni} allowed a detailed
analysis of the stellar and the gas kinematics, characterised by
maximum rotational velocities of $\pm$150 \kms\ and $\pm$175
\kms\ respectively, reached at r=10\arcsec. The central velocity dispersion 
for stars is 150 \kms, whereas gas lines have instrumental width
($<$100 \kms). The galaxy contains a source of infrared emission
detected by IRAS, and at X-ray wavelengths the published work reports
an upper limit (see Table 1). 

Our J=2--1 and J=1--0 spectra both show a weak narrow line of $\sim$30
\kms\ FWHM, centered at $\sim$2850 \kms, close to the optical redshift
of 2924 $\pm$ 66 \kms\ reported in RC3 \citep{rc3}. These results are, however, 
at odds with that quoted by \citet{dacosta} (3053 $\pm$20\,kms$^{-1}$), who 
estimated v$_{sys}$ as a weighted average between gas emission and stellar absorption 
data. If the \citet{dacosta} value is more accurate, as indicated by the 
additional spectra of \citet{bettoni}, the reported difference between the CO 
peak and the optical systemic velocity could be explained with an asymmetry in 
the distribution of cold gas. The derived molecular gas mass, $\sim$ 6\,10$^7$ 
\Msun, could be well accounted for if the emission observed came from a few 
Giant Molecular Associations (GMAs) in the center of the galaxy, as observed 
in IC\,1459. The narrowness of lines in both transitions supports this 
scenario.

\subsection{NGC 7079}

NGC\, 7079 is a weakly barred SB0 galaxy, a member of an interacting pair. 
\citet{bettoni97} detected a counterrotating disk-like structure of ionized 
gas which extends up to a radius of r=2 kpc. The radial velocities for the gas 
span from 2600 \kms\ to 2800 \kms. The stellar kinematics is typical of an 
undisturbed disk. The measured radial velocities (up to r=4.5 kpc) range from 
v$\sim$2500 \kms\ to v$\sim$2900 \kms\ and give a v$_{sys}$=2680 \kms\ and a 
central velocity dispersion of 150 \kms. No X-ray emission has been detected 
from this galaxy; the IRAS satellite detected infrared emission.

The J=2--1 and J=1--0 lines of CO are detected in this galaxy, showing the
presence of molecular gas (see Fig.1). The line profiles, $\sim$170 \kms\ 
wide at zero power, are centered on the galaxy systemic velocity, derived from
optical data. The linewidths of both CO lines agree satisfactorily
with the velocity range measured for the counterrotating ionized gas. In
contrast, the CO widths are much smaller than the velocity interval measured 
in the stellar lines. This may indicate that H$_2$ gas
is confined to the inner portion of the galaxy and that it shares the same
kinematics as the counterrotating ionized gas.
The inferred molecular gas content is M$_{mol}\sim$1.2\,10$^{8}$\Msun
up to r=2 kpc. 

\section{The enlarged sample of accreting galaxies}

The newly acquired data described above would nevertheless be
insufficent to extrapolate estimates on the global gas content to all
counterrotators. In order to improve the SEST sample on statistical
grounds we added the published data from those counterrotators
\citep{gall96,kannappan} with an estimate from any of the different
gas tracers: M$_{mol}$, M$_{HI}$, M$_{dust}$ or M$_X$.  Masses are
derived with the same assumptions used for the SEST sample. The
properties of the enlarged sample are in first part of Table 1, 
together with the list of references relevant for this
compilation.  This new sample allows a complete study of the global
gas content of counterrotators, using different gas tracers in a
statistically significant sample of 58 objects.

In order to compare counterrotators and polar rings, the available
data on a sample of 46 polar ring galaxies have also been compiled(see
references in Table 1). Data include 36 polar ring
lenticulars and spirals (\citet{rubin3,whitmore} and this work), and
10 polar ring ellipticals, known in the literature as ellipticals with
minor-axis dust-lanes, such as NGC 5128 \citep{bg}. Polar ring
ellipticals have traditionally been classified as S0s due to the
presence of a dark ring or disk in optical pictures. However, the
luminosity profiles of these galaxies do not follow an exponential
law, typical of stellar disks, but rather a r$^{1/4}$ law,
characteristic of spheroidal systems. We have therefore re-classified
these galaxies as ellipticals, whenever they appear as S0s in
catalogs. 
Not all polar ring lenticulars and spirals present in \citet{whitmore}'s
catalogue were finally included in our list. We only selected those systems where the 
perpendicularity between the ring and the stellar body is clearly visible in the 
images, discarding all systems appearing doubtful in our inspection of the catalogue. 
Altogether the sample of accreting systems includes 104 objects.

\begin{figure*}
\resizebox{\hsize}{!}{\includegraphics{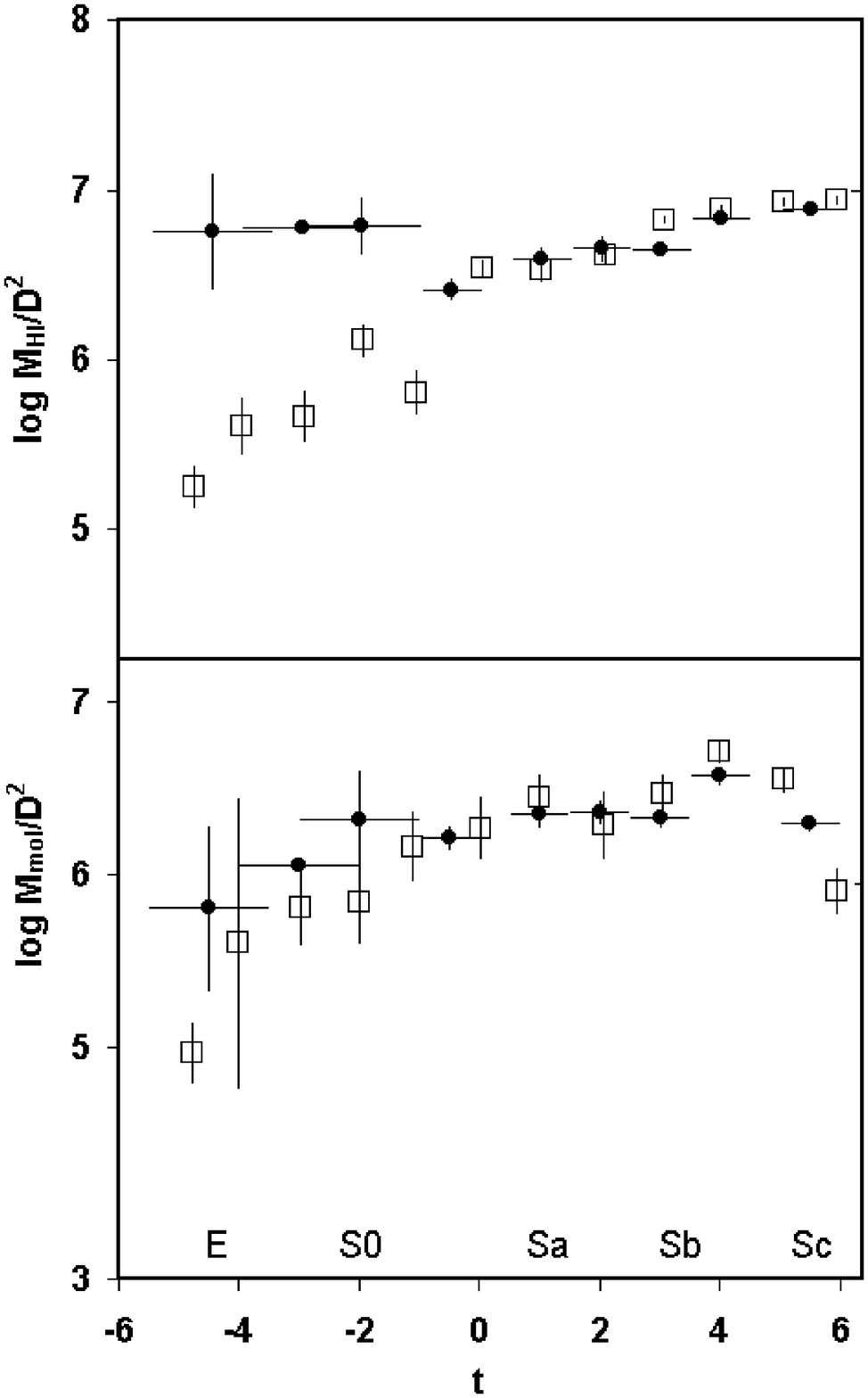}\includegraphics{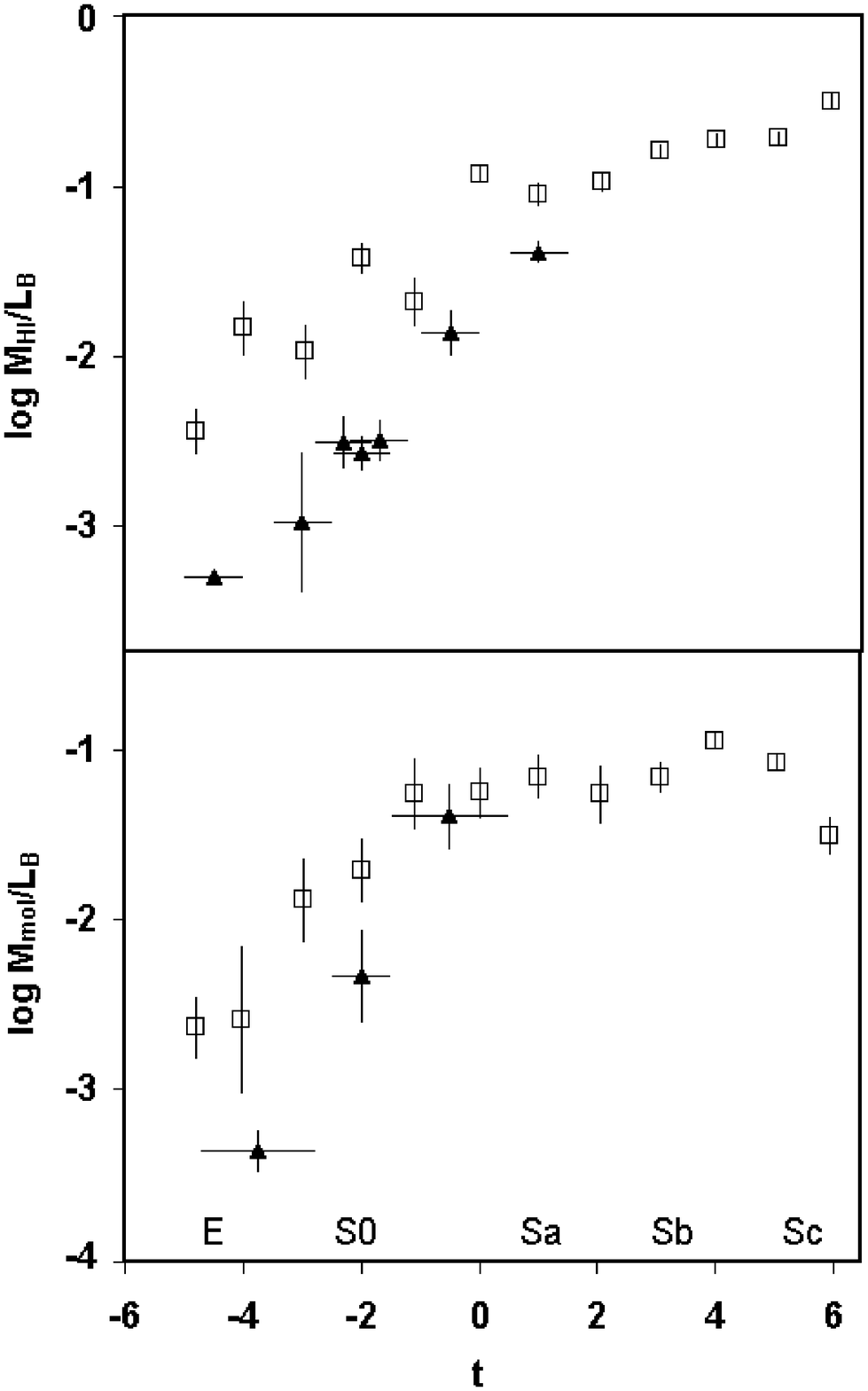}}
\caption{(Left): Log M/D$^2$ ratios in normal galaxies derived for the atomic and 
molecular gas, from the samples of \citet{casoli} (filled squares) and 
\citet{paper2} (open symbols).(Right): Log M/L ratios in normal galaxies for 
the atomic and molecular gas, from \citet{bregman} (filled squares) and 
\citet{paper2} (open symbols).}
\label{normal}
\end{figure*}

\section{Building up a comparison sample}

The main aim of this paper is to study the molecular gas content of
accreting galaxies (counterrotators and polar rings) and compare it
with the {\it average} value for {\it normal} non-interacting galaxies
as a function of the Hubble type. The first non-obvious task is the
definition of a comparison sample of {\it normal} galaxies. The sample
should contain a statistically significant number of objects. This
requirement is critical for early-type galaxies, as the majority of
accreting systems are of types earlier than Sa (morphological type
code t=1). Moreover, the sample should avoid the inclusion of abnormal
objects, suspected to be interacting and/or merging galaxies, e.g
those reported in Arp's catalogs.
      
In the past, two different research groups have built up comparison
samples in order to study the variation of the gas content of galaxies
along the Hubble sequence:
\citet{bregman} and \citet{casoli}. \citet{bregman} derived the content of 
molecular gas, HI, X-ray emitting gas and dust, working on a sample of
467 early-type galaxies, ranging from pure ellipticals (E, t=--5) to
early spirals (Sa, t=1). In their analysis they favoured the use of
the total blue luminosity (L$_B$) of galaxies as the necessary
normalisation factor, i.e., the inferred numbers being
M$_{mol}$/L$_B$, M$_{HI}$/L$_B$, M$_{dust}$/L$_B$ and M$_X$/L$_B$.
They concluded, first, that the gas content of elliptical galaxies is
dominated by the hot phase(M$_{X}>$M$_{mol}$+M$_{HI}$) and secondly,
that M$_{mol}$/L$_B$ and M$_{HI}$/L$_B$ both show a strong positive
gradient from E to Sa-type systems.  However, the number of early type
galaxies detected either in H$_2$ or in HI gas is scarce: for H$_2$, 1
E-type detected with 11 upper limits (hereafter UL) and 6 SOs detected
with 18 UL. Poor statistics cast some doubt on their conclusions.

\citet{casoli} discussed the molecular and atomic gas content for a set of 582 
objects, mainly disk galaxies, normalizing the gas masses by
D$_{25}^2$. For comparison, numbers for H$_2$ are: 3 E detected and 7
SOs detected (with 2 UL).  Results from \citet{casoli} are noticeably
at odds with that of \citet{bregman}: the sharp increase of gas
content from t=--5 to t=1 reported by \citet{bregman} (1--2 orders of
magnitude) is not confirmed using \citet{casoli} data.  This
discrepancy is illustrated in Figure \ref{normal} where the mean
values are represented as a function of t for the two samples. To
reconcile these discrepant trends one must assume an unrealistic
decrease by 2 orders of magnitude in L$_B$/D$_{25}^2$, going from E to
Sa systems. Results from both samples for types t$<$1 are however
dubious, considering the poor statistics in this range. Moreover, the
two samples include a non-negligible percentage of interacting
galaxies (estimated to be close to $\sim$20$\%$ in both
samples). Interacting galaxies should be discarded when putting
together a comparison sample of {\it normal} non-interacting objects.

These limitations were the motivation to build up a new comparison
sample of normal galaxies. Paper II by \citet{paper2} will discuss
extensively the details of this compilation. The over-all numbers give
a grand total of 1773 normal galaxies selected from a processed sample
of 3800 objects. \citet{paper2} purposedly excluded from the selected
sample those galaxies belonging to the interacting or disturbed
categories (most of them appearing in \citet{arp},
\citet{vv}, and \citet{am} catalogues). Galaxies listed in the \citet{veron} 
catalogue of AGN systems have also been excluded because in some cases their 
peculiar activity has been attributed to gas accretion. We have 
taken from \citet{paper2} the normalized values M/L$_B$ for the molecular, 
atomic and X-ray emitting gas, as well as for the warm dust content inferred from
IRAS. The global statistics for detections (and UL) are: 247 in H$_2$ (113 UL), 
774 in HI (149 UL), 196 in X-rays (661 UL) and 861 in IR (555 UL). This 
sample improves the statistics for early type galaxies compared 
to previous works, the numbers for H$_2$ being: 10 E-type detected (plus 18 
UL) and 10 lenticulars detected (plus 17 UL).  Note that the galaxies used
to build the mean values for the different ISM tracers are not always the same; 
however, the majority of galaxies in our sample (1135) have detections or upper 
limits in at least two wavebands. 

We have applied a survival analysis method to the different ensembles
of M/L$_B$ data.  This analysis tool takes properly into account both
detections and UL in order to derive representative averages. The mean
values are derived and plotted under the label `normal galaxies', and
are binned according to the morphological type code (with
$\Delta$t=1). Most noticeably, UL lower significantly the estimated
mean M/L$_B$ values for normal galaxies of early types (see
\citet{paper2}). The derivation of mean values of molecular gas content
from \citet{casoli} used survival analysis also. For HI data, all the galaxies 
of their sample were detected and so no survival analysis
needs be applied. Also, the analysis of \citet{bregman} takes into account 
the different detection rates of the various morphological types, but using
different non-parametric tests, based on rank. 

At first sight, the comparison between the mean values derived from
these three different studies \citep{bregman,casoli,paper2} shows that
Log(M$_{HI}$/D$_{25}^2$) and Log(M$_{mol}$/D$_{25}^2$) of
\citet{paper2} are intermediate between \citet{casoli} and
\citet{bregman} values.  Although the cold gas content increases by a
factor $\sim$10 from E to Sa-types, this gradient is less steep than
that reported by \citet{bregman} (see Fig. \ref{normal}).

To identify any potential bias in the galaxy samples compared in this
work (normal galaxies, counterrotators and polar rings), we have
analysed the statistical distribution of the following intrinsic
properties: M$_B$, D$_{25}$ and FIR flux (given by
m$_{FIR}$). Kolgomorov-Smirnov tests applied to these quantities
indicate that the distributions of M$_B$, D$_{25}$ and m$_{FIR}$ are
not significantly different for the 3 samples, at a confidence level
better than 95$\%$. 

\begin{figure}
\resizebox{\hsize}{!}{\includegraphics{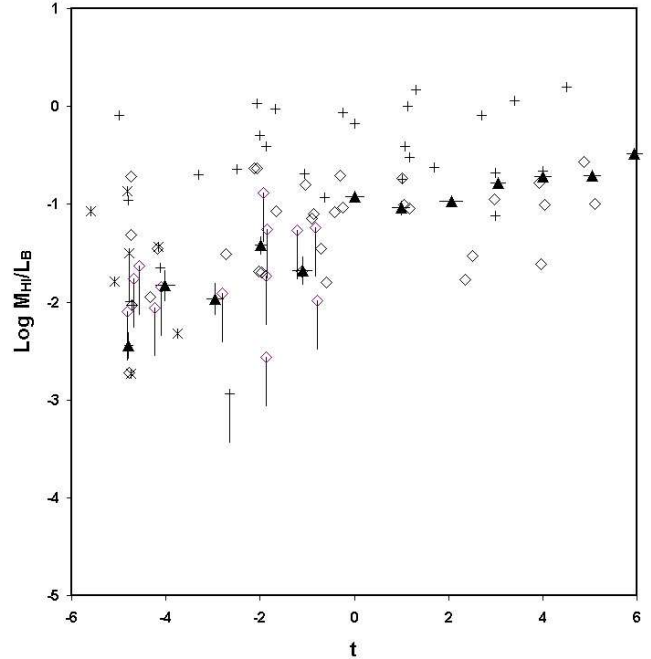}}
\caption{Plot of Log M$_{HI}$/L$_B$ versus the morphological type for the 
samples of counterrotating galaxies (open romboids), polar ring S0s and spirals 
(crosses) and polar ring ellipticals (asterisks). Filled triangles represent the reference 
values derived from the comparison sample of \citet{paper2}.}
\label{HI}
\end{figure}

\section{The ISM of gas accretors}

We discuss in this section the results obtained from the comparison of
the gas/dust content of gas-accreting and normal galaxies.  Mean
values of log M/L were obtained for each morphological type, using a
$\Delta$t=1 code binning. We studied the deviations from the reference
values issued from the survival analysis method applied to normal
galaxies (see above). The statistical significance of any difference
found between the samples was evaluated by a Student t-test applied to
the mean of the values of Log M/L, binned according to morphological type.
Fig.\ref{HI}--\ref{X} illustrate this comparison, whose
results can be summarised as follows:

\begin{itemize}

\item

Log(M$_{HI}$/L$_B$) and Log(M$_{mol}$)/L$_B$) show a large dispersion for 
accretors of types t$<$0 (Figures \ref{HI} and \ref{mol}). This result holds for 
counterrotators and polar rings. The values of the gas content for normal galaxies 
are also highly dispersed for t$<$0.

\begin{figure}
\resizebox{\hsize}{!}{\includegraphics{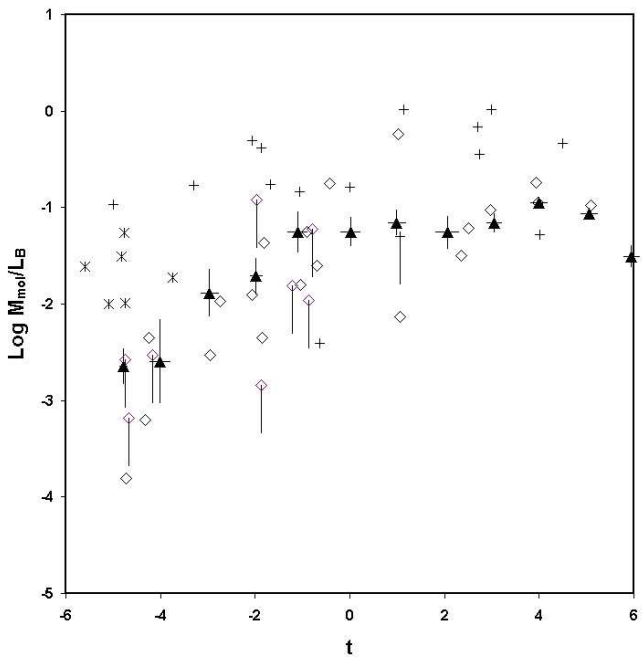}}
\caption{Variation of the molecular gas masses with the morphological type for the 
samples of counterrotating galaxies (open romboids), polar ring S0s
and spirals (crosses) and polar ring ellipticals (asterisks).  Filled
triangles represent the reference values derived from the comparison
sample of \citet{paper2}.}
\label{mol}
\end{figure}

\begin{figure}
\resizebox{\hsize}{!}{\includegraphics{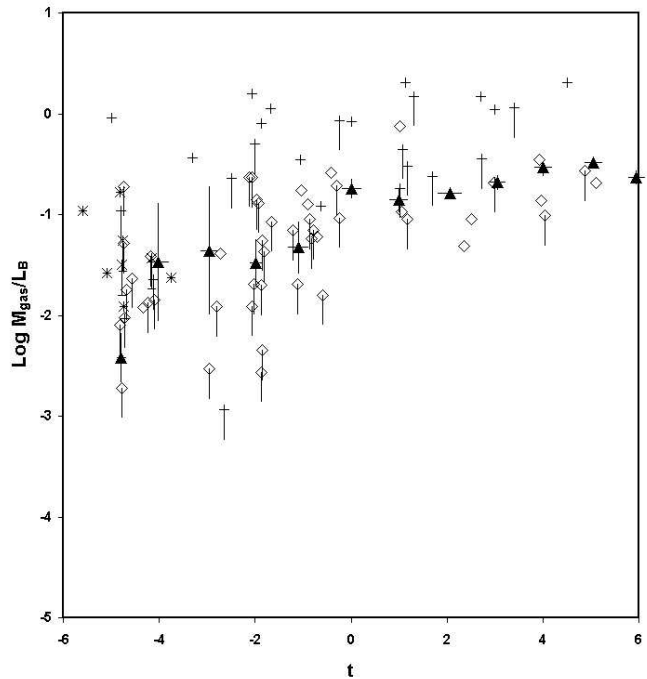}}
\caption{Change of the cold gas content (molecular and atomic)with respect to 
morphological type for the samples of counterrotating galaxies (open romboids),
polar ring S0s and spirals (crosses) and polar ring ellipticals
(asterisks).  Filled triangles represent the reference values derived
from the comparison sample of \citet{paper2}.}
\label{gas}
\end{figure}

\begin{figure}
\resizebox{\hsize}{!}{\includegraphics{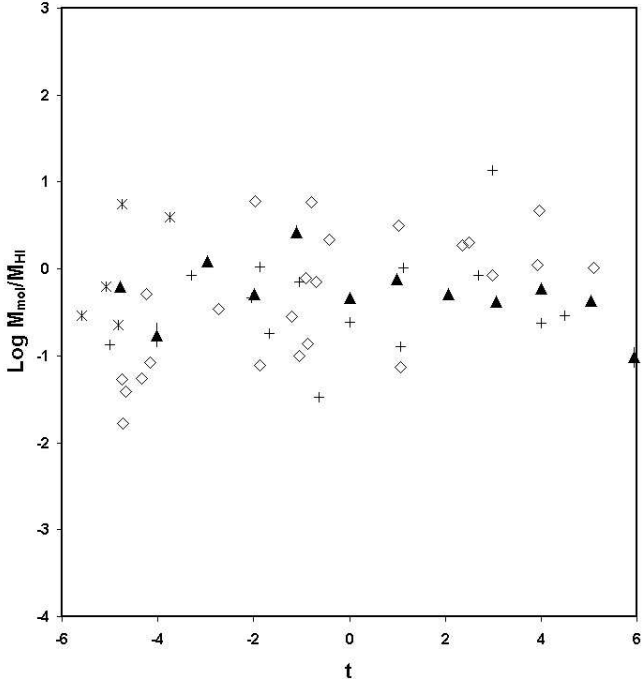}}
\caption{The ratio of molecular to atomic gas versus the morphological type. 
Symbols are: counterrotating galaxies (open romboids), polar ring S0s
and spirals (crosses), polar ring ellipticals (asterisks) and normal galaxies
(filled triangles).}
\label{mol_HI}
\end{figure}

\begin{figure}
\resizebox{\hsize}{!}{\includegraphics{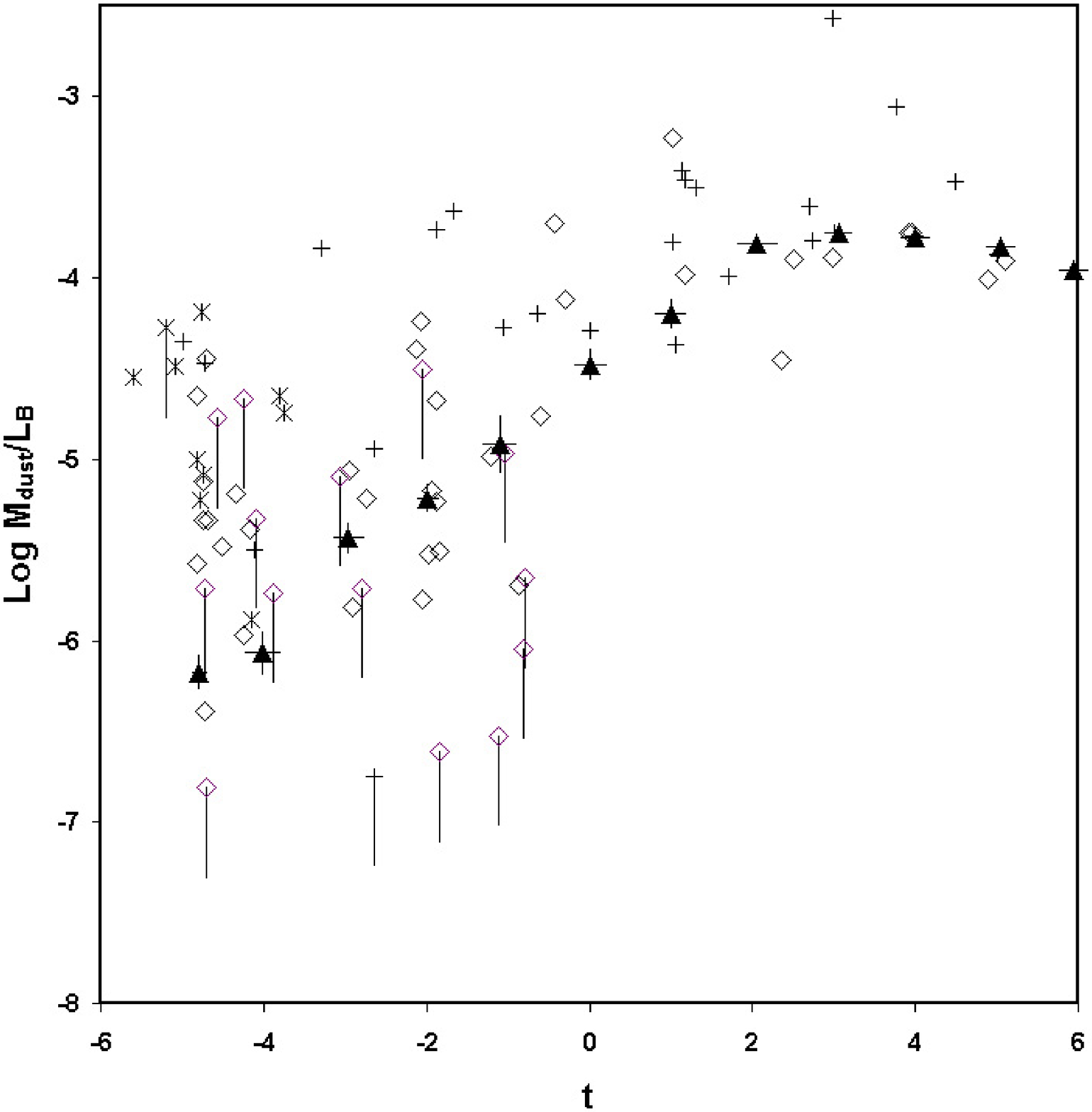}}
\caption{ Variation of the warm dust content with respect to morphological type 
for the samples of counterrotating galaxies (open romboids), polar ring S0s
and spirals (crosses) and polar ring ellipticals (asterisks).  Filled
squares represent the reference values derived from the comparison
sample of \citet{paper2}.}
\label{dust}
\end{figure}

\item Polar ring spirals and lenticulars have a HI content $\sim$1 order of
magnitude higher than normal galaxies (Fig.\ref{HI}). The reported 
difference between the samples is established with a 
99$\%$ statistical significance. In contrast, there is no significant
difference between normal galaxies and polar ring ellipticals
regarding the HI content.  This result, previously found and discussed
by \citet{richter94} and \citet{huchtmeier97} is still valid when
Log(M$_{HI}$/D$^2$), instead of Log(M$_{HI}$/L$_B$), is used as the
gas content estimator. Therefore, this tendency cannot be attributed
to any colour bias unexpectedly affecting PRs.

On the other hand, the HI content of counterrotating galaxies shows no significant 
departure from the expected normal value (Fig. \ref{HI}).

\item

The molecular gas content of polar ring galaxies lies above {\it
normal} values for all galaxy types (Fig. \ref{mol}). On average, it
is calculated that M$_{mol}$/L$_B$ is $\sim$1 order of magnitude
higher in polar rings; this result, which agrees satisfactorily with
the conclusions of \citet{gall97},  is established with a 98$\%$
statistical significance.  Therefore, the global content of cold gas
(M$_{gas}$) in S0/S-polar rings lies $>$1 order of magnitude above
standard values with a 99$\%$ certainty (see Fig. \ref{gas}). 
On the other hand, the M(mol)/M$_{HI}$ ratio in polar rings stays close to
normal values for all types (Fig. \ref{mol_HI}).

The molecular gas content of counterrotating galaxies is marginally
lower than normal for types t$<$0 (Fig. \ref{mol}).  This deficiency
is not firmly established, its statistical significance being low
(78$\%$). In contrast, counterrotating galaxies reach normal H$_2$
masses for t$>$0. Contrary to polar rings, the global content of cold
gas in counterrotators, given by Log(M$_{gas}$/L$_B$), shows no
relevant departures from normal values for all types (Fig. \ref{gas}).

However, the HI phase seems to dominate in early type counterrotators
(t$<$0).  We tentatively identify an increase by $\sim$1 order of
magnitude in the M(H$_2$)/M$_{HI}$ ratio from t=--6 to t=6, a factor
significantly larger than that observed in normal galaxies
(Fig. \ref{mol_HI}). The latter indicates that some mechanism
favouring the transformation of HI into H$_2$ might be at work for
counterrotators on the right-side of the Hubble sequence (see below).

\item

As derived from Log(M$_{dust}$)/L$_B$, polar ring galaxies have a dust 
content $\sim$0.5-1 order of magnitude higher than normal. This result is established
with a 99$\%$ statistical significance for spirals and lenticulars, whereas it is only
at a 90$\%$ certainty level for ellipticals. In contrast, galaxies with 
counterrotation have a warm dust content not significantly different to normal galaxies 
for all types (Fig. \ref{dust}).

\begin{figure}
\resizebox{\hsize}{!}{\includegraphics{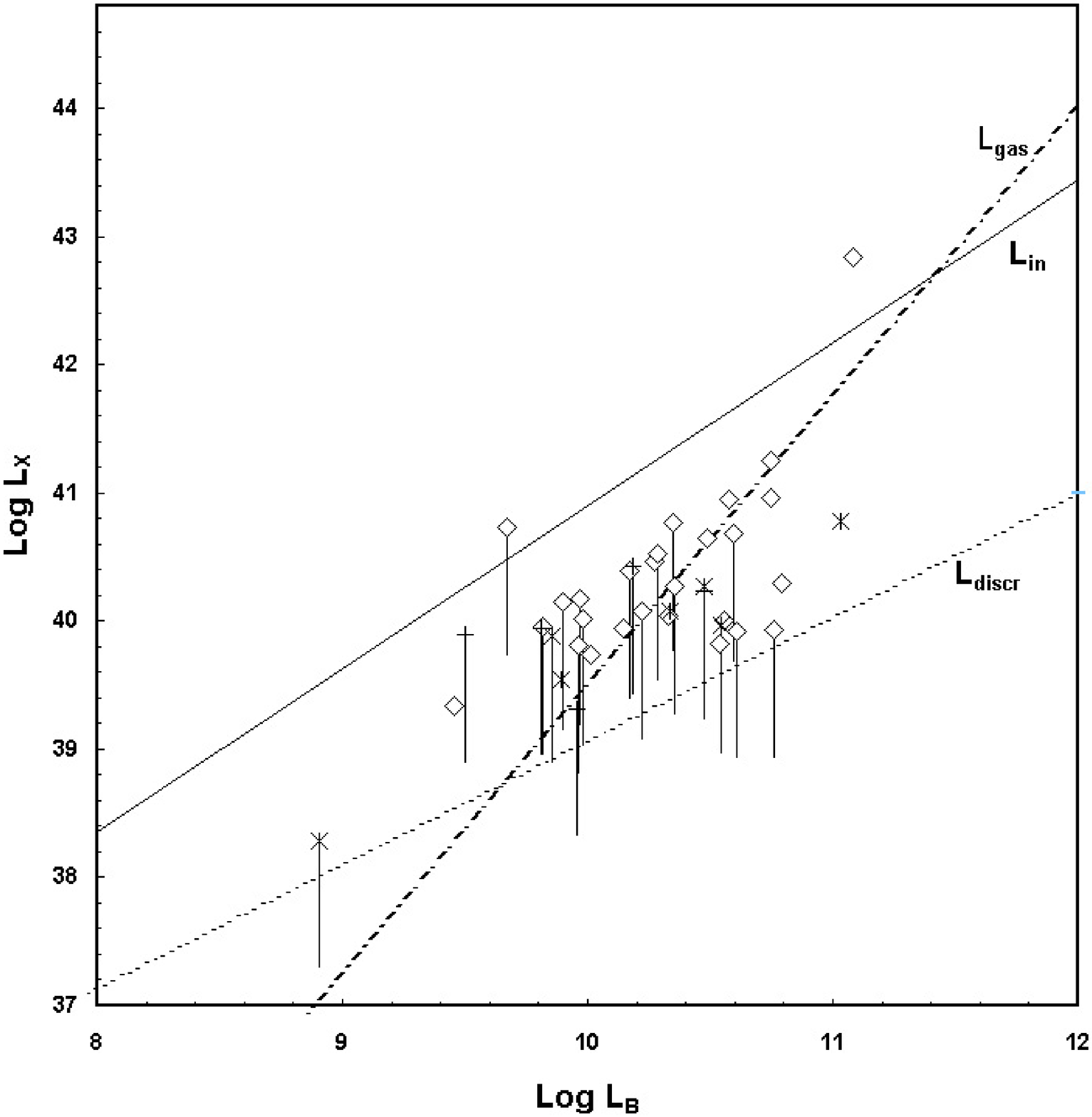}}
\caption{Plot of Log L$_{X}$  vs Log L$_B$ for the accreting 
galaxies. Symbols are as in Fig. \ref{HI}. The observed values are compared 
with different models including various X-ray luminosity sources: hot diffuse 
gas (dashed-dotted line labeled L$_{gas}$, discussed by \citet{beuing}), 
discrete sources (dashed line, labeled L$_{discr}$, discussed by 
\citet{ciotti}). All galaxies, except for NGC 4073, show luminosities lower 
than expected from steady-state cooling flow models, assuming the recent SN I 
rate 0.18 \citep{cappellaro}. }
\label{X}
\end{figure}

\item 

One fifth of counterrotators and nearly half of the polar ring
ellipticals have been detected in X-rays. On average, the estimated
masses of hot gas (see Table 1) are slightly lower than
the ones of normal galaxies.  Moreover, the slope fitted to the 17
detections in the L$_X$--L$_B$ diagram ($\sim$ 1.7, Fig. \ref{X}) lies
between the prototypical value for emission mainly due to hot diffuse
gas (L$_{gas}$, with slope $\sim$2), and the one for emission being
dominated by discrete sources(L$_{discr}$, with $\sim$1; see
\citep{ciotti} for details).  Furthermore, L$_{X}$ values for
accretors are below the emission level predicted by cooling flows
models, assuming the recent SN I rate equal to 0.18 (i.e., 1.8
10$^{-3}$ SN I events per year per unit of 10$^{10}$ L$_B/L_{\sun}$)
\citep{cappellaro}. We can conclude that the observed normalised X-ray 
luminosity of accretors needs no huge starburst event as an explanation.

\item
A source of uncertainty for M$_{mol}$ comes from the undersampling of
some of the sources. Apart from NGC\,3497, only single point maps
centered in the galaxy nuclei are available from the SEST
observations. Therefore, the derived M$_{mol}$ should be taken as a
lower limit for the SEST sample, as well as for some of the galaxies
in the enlarged sample of accretors for which no complete maps are
available.  Although it is difficult to evaluate accurately the bias
introduced on M$_{mol}$, we are confident to have on average
$\sim$70$\%$ of the total molecular gas masses under the SEST beam for
spirals. This estimate is based on the comparison between the mean
ratios of D$_{beam}$/D$_{25}$ ($\sim$0.4, where
D$_{beam}$=22$\arcsec$) for the SEST galaxies and the predicted ratios
of D$_{CO}$/D$_{25}\sim$0.5 for typical spiral galaxies. The values of
D$_{CO}$, defined as the diameter of the canonical distribution of CO
in Virgo spirals which contains 70$\%$ of the total CO flux
(\citet{kenney88}), indicate that molecular gas is highly concentrated
in the inner optical disks. Therefore, the reported differences (a
factor of $\sim$10!) in the molecular gas content between polar rings
and counterrotators can hardly be attributed to a systematic
undersampling of counterrotators, considering also that undersampling in 
CO maps affects polar ring galaxies to a comparable extent.
Furthermore, results obtained using HI as a tracer of cold gas (much less affected by 
undersampling), confirm a similar trend as that shown by CO: polar rings have a gas
content significantly higher than counterrotators. 
\end{itemize}

\section{Discussion and conclusions}

Gas and stars along polar orbits can be explained as the result of the
acquisition of cold infalling gas by an accreting galaxy. The accreted
gas can smear out into a ring after a few orbital periods. The fate of
this ring will depend on its orientation, relative to the mass
distribution of the accretor, and most importantly, on the
mass/self-gravity of the gas. A polar ring may form after a high angle
impact with gas which has a spin perpendicular to the equatorial plane
of the accretor, remaining in an equilibrium configuration for several
Hubble times. In contrast, counterrotating galaxies may form after a
low angle impact with a gas disk which has a spin antiparallel to that
of the accreting system.

In the most general case, however, the impact angle is intermediate between 
polar and planar. In this case the orbits of gas clouds will
experience differential precession in the non-spherical potential of the 
galaxy, characterized by a quadrupole component \citep{steiman82}. The latter 
applies for disk galaxies (axisymmetric) and ellipticals (triaxials). 
\citet{sparke86} and \citet{arnaboldi94} have studied in detail the dynamics 
of self-gravitating annuli of matter inclined to the principal axes of 
axisymmetric and triaxial potentials. If the strength of gas self-gravity is 
negligible, the inclined ring may rapidly settle towards the equatorial plane, 
appearing either as a {\it co}- or as a {\it counter}-rotating disk. In  
contrast, if the gas ring is {\it heavy}, the self-coupling can stabilise the 
ring for several Hubble times. In an intermediate case, several subrings 
(near polar or close to the equator), characterized by different precessing 
rates, can coexist in a single galaxy (e.g. NGC 660).

Studying the morphology of polar rings, \citet{vangorkom} and
\citet{sage1} have found that their ages may vary from 400 Myr to
$\ge$ 4 Gyr, if the smooth rings are the oldest. Some polar rings or
inclined rings could be the result of recent acquisitions, whereas
others appear to be evolved systems.  However, twisted polar rings are
uncommon and some have had time to form stars.  This requires the
existence of a stabilising mechanism. The observations of atomic and
molecular gas show that the quantity of gas mass present in polar
rings is sufficient to stabilise them through self-gravity
(\citet{vangorkom, sage1, gall97}, and {\it this work}). We derived a
global content of cold gas (M$_{gas}$) in polar rings which is 1--2
orders of magnitude higher than in normal galaxies.

In contrast to polar rings, the derived content of cold gas in
counterrotators is close to normal.  Although counterrotators and
polar rings probably share a common origin, the estimated gas masses
confirm that {\it light} gas rings may have evolved faster. If the
mass of gas originally accreted is not sufficient to stabilise the
ring through self-gravity, the ring settles toward the equatorial
plane in less than a Hubble time. In this case, the merger relic could
be a counterrotator.  Once the gas disk has settled to the plane with
an antiparallel spin it can interact with the gas of the primary
disk. Since the two components have opposite rotating directions,
there can be large-scale shocks and angular momentum annihilation when
they come into contact. Near this transition region the transformation
of atomic into molecular gas could be enhanced, especially if the
primordial gas content is high, i.e., for late-type accretors.
Confirming these expectations, the measured M(H$_2$)/M$_{HI}$ ratio
seems larger in counterrotators than in normal galaxies for types
t$>$0.

In the course of this process, a starburst might be triggered in the
circumnuclear molecular gas disk \citep{santi2}. The time-scale for
gas infall could be extremely short, being close to the free-fall
time, i.e., $\sim$10$^{7-8}$Myr. The mass of gas involved in the
starburst episode however is kept low enough (10$^{8-9}$M$_{\sun}$)
for a typical counterrotating galaxy. In polar rings, although the
cold gas content is larger than the one in normal galaxies, star
formation in the dynamically stable ring proceeds calmly.  Confirming
this scenario it was found that counterrotators, polar rings and
normal galaxies have a similar content of hot gas, according to their
normalized X-ray luminosities. Also, the normalized L$_{FIR}$ lies
within the typical boundaries of aged (T$\sim$1Gy) mild starbursts,
far from the values characteristic of massive mergers (see
\citet{read}).

Acknowledgments

We would like to thank dr. G. Paturel for kindly making available to
the authors the FIR raw data of LEDA database and to Dr. F. Ochsenbein 
for the changes made to the Vizier's query form after our request. 
Thanks to the referee's comments for useful suggestions on the statistical 
analysis.  This research made use of Vizier service \citep{vizier} and of 
NASA's Astrophysics Data System Abstract Service, mirrored in CDS of
Strasbourg.  SGB and ARF thank financial support from the Spanish
CICYT under grant number PB96-0104 and CICYT-PNIE under grant number
1FD1997-1442. GG has made use of funds from University of Padova
(Fondi 60\%-2000).

\end{document}